\documentstyle[aps,prl, multicol]{revtex}

\begin{document}

\draft
\preprint{HEP/123-qed}
\title{Tuning by an Electric Field of Spin Dependent\ 
Exciton-Exciton Interactions in Coupled Quantum Wells}

\author{G. Aichmayr, M. Jetter, and L. Vi\~na}
\address{
Departamento de Fisica de Materiales C-IV, Universidad Autonoma de Madrid,
Cantoblanco,
E-28049 Madrid, Spain}

\author{J. Dickerson, F. Camino, and E.E. Mendez}
\address{
Department of Physics and Astronomy, SUNY at Stony Brook, N.Y. 11794-3800, USA}
\date{\today}
\maketitle
\begin{abstract}
We have shown experimentally that an electric field decreases the energy 
separation between the two components of a dense spin-polarized exciton 
gas in a coupled double quantum well, from a maximum splitting of $\sim 4$~meV 
to zero, at a field of $\sim $35 kV/cm.  
This decrease, due to the field-induced deformation of the exciton wavefunction, 
is explained by an existing calculation of the change in the spin-dependent 
exciton-exciton interaction with the electron-hole separation. However, a new 
theory that considers the 
modification of screening with that separation is needed to account for the 
observed
dependence on excitation power of the individual energies of the two exciton 
components.
\end{abstract}

%\narrowtext
\begin{multicols}{2}
Spin dependent exciton-exciton interactions in low dimensionality semiconductors
have been studied extensively
during this decade
\cite{Damen91,Stark92,Dareys93,Amand94,Vina96,Joaquin96,Joaquin97,LeJeune98}.
The first observation of an energy splitting between the two spin polarized
components of an exciton gas without a magnetic field was reported
by Damen et al.
\cite{Damen91}. Since then this effect has been considered both experimentally
\cite{Vina96,LeJeune98} and theoretically
 \cite{Joaquin96,Joaquin97}.
Fern\'andez-Rossier {\it et al.} attributed that splitting to a density- and 
spin-dependent
modification of the two dimensional (2D) exciton binding energy ($E_{2D}$)
by two competing processes
\cite{Joaquin96,Joaquin97}.
One of these many-body processes is the exchange interaction ($I_{EC}$) between 
electrons (and holes) of  different
excitons, which always {\it increases} $E_{2D}$ in proportion to the 
spin-polarized exciton density, so that the change, $\Delta E^{\pm}_{2D}$, is
$\Delta E^{\pm}_{2D} \propto n_X^{\pm}I_{EC}$, where $n_X^{\pm}$ refers to the
spin +1 and -1 exciton populations, respectively \cite{foot1}.
The second process is the vertex correction ($I_{VC}$) to the Coulomb interaction
between electrons and holes of different excitons.  This correction effectively
reduces the electron-hole attraction due to the occupation
of final states in the electron-hole scattering processes and 
{\it lowers} $E_{2D}$ proportionally to $n_X^{\pm}I_{VC}$.
Hence, neglecting the small coupling between +1 and -1 states due to valence 
band mixing, the total change of the exciton binding energy is
$\Delta E^{\pm}_{2D} \propto n_X^{\pm}(I_{EC}-I_{VC})$.

When the overlap between the electron and hole wavefunctions is large the 
vertex
correction outweighs exchange effects (that is, $(I_{EC}-I_{VC})< 0$) and the 
exciton binding energy is reduced (with respect to that of a single exciton).  
Moreover, when $n_X^{+} > n_X^{-}$, the reduction is such that 
$|\Delta E^{+}_{2D}| > |\Delta E^{-}_{2D}|$, leading to a smaller binding 
energy of the +1 excitons than that of the -1. 
As a result an energy splitting $\delta \epsilon $ between the components of a 
dense spin-polarized quasi 2D exciton gas is expected.  Such a splitting has 
been observed in time-resolved photoluminescence (PL) experiments in quantum 
wells, in which a circularly polarized light pulse
($\sigma ^+$) creates excitons with spin +1 that gradually relax into the 
spin -1 state, until both populations are equal.  The PL spectrum 
corresponding to the spin +1 exciton has been found to be at higher energy 
than that of the -1 exciton, with the splitting between the two, 
$\delta \epsilon$, decreasing exponentially with time.

However, until now no experiment has tested the theoretical prediction of 
a decrease or even a sign reversal in $\delta\epsilon $ caused by a 
modification of the vertex correction and  the
exchange term when the overlap of electron and hole wavefunction is 
reduced \cite{Joaquin97}. 
We report here the first evidence of such a change, determined from 
measurements of $\delta\epsilon $ in coupled double quantum wells 
(CDQW) subjected to a longitudinal electric field that separates spatially 
the electrons and holes.
We not only find a good agreement between the observed energy splitting 
and the predicted one, but
our results go beyond available models. Specifically, we have found that 
screening of
these many-body corrections is strongly reduced by the field, an effect 
not contemplated by
theory so far.

The material structure used for the experiments was an n$^+$-i-p$^+$  
GaAs-AlGaAs heterojunction.  The active region consisted of ten CDQWs 
periods separated from each other by a 200~{\AA} Al$_{0.3}$Ga$_{0.7}$As 
layer, and each formed by two 50 {\AA} GaAs wells with a 20~{\AA}  
Al$_{0.3}$Ga$_{0.7}$As barrier in between.  An undoped 
Al$_{0.3}$Ga$_{0.7}$As layer on each side of the CDQW stack 
(100~nm and 80~nm thick, respectively) completed the intrinsic (i) 
region of the structure, whose p$^+$ and n$^+$ electrodes were a 
500~nm p$^+$-doped GaAs:Be
layer and a 1~$\mu m$ thick n$^+$-doped GaAs:Si layer on a [100]
GaAs substrate, respectively. 

A schematic representation of the potential profiles and energy levels of 
the CDQW structure at a field of 14~kV/cm is shown in Fig.~\ref{PLE}a, 
where $e1$ and $e2$ represent the coupled-well states in the conduction 
band, and $hh1$ and $hh2$ the corresponding states for heavy holes in the 
valence band. (For simplicity, the light hole states have not been shown.) 
The field-induced symmetry breaking of the wave functions makes all the 
transitions optically allowed and the ground-state exciton ($X_{e1hh1}$) 
becomes spatially indirect due to the
localization of electrons and holes  in wells A and B, respectively
\cite{Chen87,Mendez88,Voisin88}.

The experimental electric-field dependence of the various excitonic 
transitions, 
obtained from photoluminescence excitation (PLE) measurements on the 
above heterostructure, is summarized in Fig.\ref{PLE}b.
The lower lying indirect transitions $X_{e1hh1}$ and $X_{e1lh1}$
undergo a red shift with increasing field, primarily because of the 
tilting of the
potential, although at high fields the Stark effect contribution becomes 
appreciable \cite{Vina87}.
The two direct transitions $X_{e1hh2}$ and $X_{e2hh1}$  have a less marked 
field dependence and at high fields they approach each other in energy.
The identification of the transitions and determination of the effective 
field in the intrinsic region was done through a comparison with calculated 
values obtained by solving numerically the Schr\"odinger
equation within the envelope function approximation \cite{Bastard}. (See
dotted lines in Fig.~\ref{PLE}b.)  

For the study of the spin-dependent exciton-exciton interaction, time 
resolved PL measurements were performed with a standard up-conversion
set-up using a Ti:Saphire laser with 2~ps pulse width. For
polarization-selective excitation and PL detection, $\lambda /4$-wave plates
were placed in the paths of the laser beam and the PL collecting optics. The
sample was mounted in a cold-finger cryostat and held at $T=8$~K. 

Our focus is on the two spin components of ground-state excitons,
$X_{e1hh1}$, created after optically induced electrons relax from the $e2$ 
to the $e1$ states 
(Fig.~\ref{PLE}a). An exciton population n$_X\approx 5\times 
10^{10}$~cm$^{-2}$ was created with 15~mW pulses of circularly polarized 
light, whose photon energy was tuned to the $X_{e2hh1}$ direct exciton.  
Initially a $+1$ exciton population was created; however, spin-flipping 
processes gave rise to a non-equilibrium mixture of $+1$ and $-1$ excitons 
with populations $n_X^+$(majority) and $n_X^-$ (minority), respectively. 
The degree of polarization, defined as
$P=\frac{n_X^+-n_X^-}{n_X^++n_X^-}$, was found to be initially $P_0=0.5 
\pm 0.05$, independent of the electric field.

Also almost independent of the field was the excitonic PL rise time, which 
amounted to $110\pm 10$~ps for the $X_{e1hh1}^+$
and to $140\pm 10$~ps for $X_{e1hh1}^-$. In contrast, the PL decay time, 
$\tau_d$, increased linearly (almost independently of spin polarization) 
with increasing field, from 400~ps at ${\cal E}=0$  to 1000~ps at 
${\cal E}=35$~kV/cm. For even higher fields the decay time decreased 
again, and at ${\cal E}=40$~kV/cm it dropped to 250~ps. These two regimes 
have been explained in terms of a reduction of wave function overlap and 
of tunneling of the carriers out of the
wells, respectively \cite{Kohler88,Alexandrou90}.

A measure of the exciton-exciton interaction is the energy difference 
between the PL spectra of the X$^+_{e1hh1}$ and X$^-_{e1hh1}$ excitons, 
plotted in the insets of Fig.~\ref{splitting} for three representative 
electric fields.  These spectra were taken at a time, $t_d=32$~ps, at 
which the short-lived peak broadening of the high energy side of the PL 
spectra due to
electron-hole (e-h) plasma luminescence was already negligible \cite{foot2}.
The main body of Fig.~\ref{splitting} shows the time evolution of the peak 
energies of both spectra.  At ${\cal E}=7$~kV/cm (Fig.~\ref{splitting}a), 
the +1 (-1) exciton peak shifted to lower (higher) energies and their energy 
difference decreased exponentially with a time constant $\tau _{sd}=180$~ps. 
At higher fields, aside from a field induced ``red-shift" of both spectra, 
the time dependence of the +1 exciton peak was not affected by the field, 
while the -1 exciton time dependence was quite different: at 
${\cal E}=23$~kV/cm its energy remained constant with time while at 
${\cal E}=35$~kV/cm it evolved as the +1 exciton.

The electric-field dependence of $\delta \epsilon$, defined as the average 
splitting taken at delays $t_d=7$, 15 and 32~ps, is summarized in 
Figure~\ref{splitting(E)}. The maximum splitting of $\sim 4$~meV is reached 
at zero field, and then it decreases linearly until it becomes zero at 
${\cal E}\simeq 35$~kV/cm. The parameter $d$ in the upper scale of 
Fig.~\ref{splitting(E)} represents the average separation (along the 
growth direction) between 
electron and hole as calculated from the field dependent expectation values. 
The zero splitting is reached at an e-h separation of 
$\sim 60$~{\AA}, which is a factor of 2.5 larger than the value predicted 
by the model 
\cite{Joaquin97}. This discrepancy is not surprising in view of the 
simplifications of 
the model, such as zero temperature and a strictly-2D confinement.

Since the splitting ($\delta \epsilon =E_X^+-E_X^-\propto 
n_XP(I_{VC}-I_{EC})$) is proportional 
to the total exciton population, the polarization, and the many-body 
corrections, and since $P$ is not much affected by the field, at a given 
excitation power any change of $\delta \epsilon $ with field has to be 
related to 
a modification of the exciton-exciton interaction. Furthermore, $\delta 
\epsilon (t)$ and 
$P(t)$ decay with the same time constant, which verifies the expected 
correlation
between $n_X^+-n_X^-$ and $\delta \epsilon $ assuming a constant $n_X$ 
(i.~e., that the 
exciton lifetime is much longer than the other decay times).

An analysis of the power dependence of the $\pm $1 exciton's energies at a
short delay time, and hence at constant polarization, allows a direct
comparison of the experiments with theory and a test of its
underlying assumptions. The energy positions of the two spin polarizations
are displayed in Fig.~\ref{density} as a function of excitation power. Let us
consider first the energy splitting between the two components. At low fields 
(Fig.~\ref{density}a) $\delta
\epsilon $ increases considerably with power, in agreement with previous 
experimental 
results at zero field \cite{Vina96} and the predictions of theory 
\cite{Joaquin97}. 
At intermediate fields (${\cal E}=23$~kV/cm, 
Fig.~\ref{density}b), the
splitting still grows with increasing power, however the values of $\delta
\epsilon $ are reduced with respect to the low field case, as a consequence
of the enhanced e-h separation. At a  field of ${\cal E}=35$~kV/cm,
when the splitting vanishes, it remains zero at all excitation powers, in
agreement with the theoretical model, which predicts that at a certain $d$
the $I_{VC}$ and $I_{EC}$ corrections become equal, independently of the
total exciton population.

Turning our attention to the individual energies, we have observed that 
the power dependence of the individual energies of the two spin
components has a strong field dependence, as seen in Fig.~\ref{density}. 
At low
fields, the energy of the +1 excitons does not change appreciably  with
power, while that of the -1 excitons decreases markedly. A similar behavior
observed in zero field measurements \cite{Vina96} could be explained only 
after including
screening corrections to the VC and EC effects. Without these corrections, 
the theory
predicted positive slopes in the power dependence of both $E_X^{+}$ and $E
_X^{-}$, in contradiction with the experiments. (Including the effects of 
screening was not
necessary to explain the splitting since screening depends only on $n_X$ 
and not on the 
individual $n_X^{\pm }$.) 
The same situation is observed in the present experiment at 7~kV/cm: 
the predicted blue
shift of $E_X^{\pm }$, due to the negative correction to the binding
energies $\Delta E_{2D}^{\pm }\propto n_X^{\pm }(I_{EC}-I_{VC})$, is
cancelled and overcompensated by the screening for $E_X^{+}$ and 
$E_X^{-}$, respectively.

At intermediate fields (Fig.~\ref{density}b) both PL components shift to 
higher energy with increasing
excitation power, although at a different rate. Finally, at the highest fields 
(Fig.~\ref{density}c) 
the rate of ``blue-shift" is the same for both components, and the splitting 
remains zero at 
all powers. The latter behavior for $E_X^{+}$ and $E_X^{-}$ is similar to 
that found by a
model whose primary focus was on the energy splitting and therefore did not 
include screening \cite{Joaquin97}.
At first sight, such an agreement at high fields may seem fortuitous, but 
since 
at intermediate fields the behavior found experimentally was intermediate 
between the two 
extreme fields, it is reasonable to suggest that screening decreases with 
increasing
e-h separation and eventually vanishes. This reduction of screening is far 
from intuitive; its explanation will require a theoretical model that 
takes screening properly
into account.

Experimental limitations prevented us from testing the prediction of a 
negative splitting (ferromagnetic phase) when the electron-hole separation 
is sufficiently large \cite{Joaquin97}.   Field-induced tunneling of 
carriers out of the wells, and a concomitant exponential increase in 
photocurrent, limited the
maximum applicable field to $\sim 35$~kV/cm, which probably falls short 
of the 
fields needed to explore the complete phase diagram \cite{Joaquin97}.

In conclusion, we have demonstrated that it is possible to alter spin-dependent
exciton-exciton interactions in a GaAs CDQW-structure by applying an external
electric field.
This effect manifests itself as a change in the energy splitting between the 
two components 
of a dense spin-polarized exciton gas. The
splitting can be tuned from a maximum value of 4~meV at zero field to 0~meV at
${\cal E}=35$~kV/cm, which corresponds to an e-h separation of $\sim 60$~{\AA}.
The electric field dependence of the two competing processes, inter-excitonic
exchange and vertex correction, reduces the splitting, due to
a decrease (increase) of the vertex (exchange) correction with increasing
electron-hole separation. Although the applicability of the
model is limited, reasonable agreement is achieved
between experiment and theory. 
Going beyond existing models, our observations indicate that screening of 
the exciton-exciton
interaction is also a function of the separation of electrons and holes, 
a finding
we hope will stimulate further theoretical treatments.

\acknowledgments
We thank J.~Fern\'andez-Rossier and C.~Tejedor for useful discussions. 
We acknowledge R.~Ruf's contribution to the epitaxial growth of the 
heterostructures used
in this study. This work has
been partially supported by the Fulbright Commission, the 
Fundaci\'on Ram\'on Areces,
the Spanish DGICYT(PB96-0085) and the US Army Research 
Office. G.~A.~thanks the Austrian BMWV.

\end{multicols}
\pagebreak

\begin{figure}
\caption{(a) Schematic structure and levels of the coupled double quantum well.
The solid, dotted and dashed arrows indicate the excitation, relaxation and
recombination processes, respectively. The solid (dotted) line is the symmetric
(antisymmetric) wave function calculated for ${\cal E}$=14~kV/cm. The light
hole levels are not shown, for simplicity.
(b) Excitonic energies measured by PLE spectroscopy at $T=8$~K as a function of
electric field (points). The upper scale shows the applied voltage 
corresponding to the internal field.
The lines represent the calculated energies (see text for details).}
\label{PLE}
\end{figure}

\begin{figure}
\caption{Peak positions of the two spin polarized exciton components as
a function of time for three different electric fields,
at 15mW excitation power.
The excitation is $\sigma ^+$ polarized and set to the $X_{e1hh2}$ energy.
The solid lines are guides to the eye. The line for the +1 exciton is 
always the
same, only shifted rigidly in energy by 6meV (b) and 13meV (c) with 
respect to
the position in (a).
The insets show the normalized spectra for $\sigma ^+$ (solid dots) 
and $\sigma ^-$
(open dots) polarization of the emission, 32~ps after excitation.
The arrows mark the peak positions. All measurements were taken at 
$T=8$~K.}
\label{splitting}
\end{figure}

\begin{figure}
\caption{Energy splitting between the +1 and -1 exciton components 
as a function
of electric field. Each data point corresponds to an average over 
three measurements at
delay times $t_d=7$, 15 and 32~ps. The upper axis represents the 
electron-hole separation along the growth direction (z) defined as 
$d=<\psi _{electron}|z|\psi _{electron}>-<\psi _{hole}|z|\psi _{hole}>$.
The excitation power was 15~mW and the sample was
held at 8~K. The solid line is a linear fit to the data.}
\label{splitting(E)}
\end{figure}

\begin{figure}
\caption{Position of the $E_X^+$ and $E_X^-$ PL peaks for three
different electric fields as a function of excitation power. 
Excitation was set to the
$X_{e2hh1}$ energy and was $\sigma ^+$ polarized. The solid lines 
are guides to the
eye. All measurements were performed at $T=8$~K and $t_d=15$~ps.}
\label{density}
\end{figure}


\begin{thebibliography}{10}
\bibitem{foot1}
Throughout this work only excitons with spin $\pm 1$, which are the only ones
  which couple to light, will be considered.
\bibitem{foot2}
This contribution to the emission was taken into account to fit the excitonic
  PL peaks at short times ($t_d \le 15$~ps). It resulted from the simultaneou
  excitation of the $X_{e2hh1}^{+1}$ and continuum states in other bands.
\end{thebibliography}
\end{document}